\documentclass[fleqn,usenatbib,letters]{mnras}

\usepackage{newtxtext,newtxmath}

\usepackage[T1]{fontenc}

\usepackage{orcidlink}

\usepackage{xspace}
\newcommand{\Ha}{\ifmmode {\mathrm{H}\alpha} \else H$\alpha$\fi\xspace}
\newcommand{\Hb}{\ifmmode {\mathrm{H}\beta} \else H$\beta$\fi\xspace}
\newcommand{\hii}{H$\,${\sc ii}\xspace}

\newcommand{\Hii}{\ifmmode \rm{H}\,\textsc{ii} \else H~{\textsc{ii}}\fi\xspace}
\newcommand{\Hi}{\ifmmode \rm{H}\,\textsc{i} \else H~{\textsc{i}}\fi\xspace}
\newcommand{\Nii}{\ifmmode [\text{N}\,II]\lambda 6584 \else [N~{II}]$\lambda 6584$\fi\xspace}
\newcommand{\nii}{\ifmmode [\text{N}\,II] \else [N~{II}]\fi\xspace}
\newcommand{\Oii}{\ifmmode [\rm{O}\,II]\lambda 3727 \else [O~{II}]$\lambda\lambda$3727,29\fi}
\newcommand{\oii}{\ifmmode [\rm{O}\,II] \else [O~{II- }]\fi}
\newcommand{\Oiii}{\ifmmode [\rm{O}\,III]\lambda 5007 \else [O~{III}]$\lambda$5007\fi}
\newcommand{\oiii}{\ifmmode [\rm{O}\,III] \else [O~{III}]\fi}

\DeclareRobustCommand{\VAN}[3]{#2}
\let\VANthebibliography\thebibliography
\def\thebibliography{\DeclareRobustCommand{\VAN}[3]{##3}\VANthebibliography}

\usepackage{graphicx}	
\usepackage{amsmath}	
\title[Short title, max. 45 characters]{Stripped and Enriched: The Role of Ram--Pressure in Shaping Chemical Enrichment of Galaxies at Intermediate Redshift}

\author[A. Khoram]{Amir H. Khoram$^{1,2,3}$\thanks{amirhossein.khoram@unibo.it}\orcidlink{https://orcid.org/0009-0009-6563-282X},
Bianca Poggianti$^{1}$\orcidlink{https://orcid.org/0000-0001-8751-8360},
Alessia Moretti$^{1}$\orcidlink{https://orcid.org/0000-0002-1688-482X},
Benedetta Vulcani$^{1}$\orcidlink{https://orcid.org/0000-0003-0980-1499},
Mario Radovich$^{1}$\orcidlink{https://orcid.org/0000-0002-3585-866X},
\newauthor
Ariel Werle$^{1}$\orcidlink{https://orcid.org/0000-0002-4382-8081},
Marco Gullieuszik$^{1}$\orcidlink{https://orcid.org/0000-0002-7296-9780},
Amirnezam Amiri$^{4}$\orcidlink{https://orcid.org/0000-0002-8553-1964},
Sirio Belli$^{2}$\orcidlink{https://orcid.org/0000-0002-5615-6018},
Letizia Bugiani$^{2,3}$,
Neven Tomicic$^{5}$\orcidlink{https://orcid.org/0000-0002-8238-9210},
\newauthor
Giorgia Peluso$^{3}$\orcidlink{https://orcid.org/0000-0001-5766-7154},
Eric Giunchi$^{2}$\orcidlink{https://orcid.org/0000-0002-3818-1746},
Johan Richard$^{6}$\orcidlink{https://orcid.org/0000-0001-5492-1049},
\\
$^{1}$INAF-Osservatorio Astronomico di Padova, vicolo dell’Osservatorio 5, 35122 Padova, Italy\\
$^{2}$Dipartimento di Fisica e Astronomia, Università di Bologna, Via Gobetti 93/2, I-40129, Bologna, Italy\\
$^{3}$INAF, Astrophysics and Space Science Observatory Bologna, Via P. Gobetti 93/3, I-40129 Bologna, Italy\\
$^{4}$Department of Physics, University of Arkansas, 226 Physics Building, 825 West Dickson Street, Fayetteville, AR 72701, USA\\
$^{5}$Department of Physics, Faculty of Science, University of Zagreb, Bijenička 32, 10 000 Zagreb, Croatia\\
$^{6}$Univ. Lyon, Univ Lyon1, ENS de Lyon, CNRS, Centre de Recherche Astrophysique de Lyon UMR5574, F-69230 Saint-Genis-Laval, France
}

\date{Accepted XXX. Received YYY; in original form ZZZ}

\pubyear{\the\year{}}

\begin{document}
\label{firstpage}
\pagerange{\pageref{firstpage}--\pageref{lastpage}}
\maketitle

\begin{abstract}
The chemical evolution of galaxies is shaped by their star formation histories and the exchange of gas with their environments. Metallicity provides key insights into these processes, reflecting the interplay between star formation and gas flows. A fundamental aspect of this evolution is the mass–metallicity relation, which captures the strong correlation between a galaxy stellar mass ($\mathrm{M_\star}$) and its gas-phase oxygen abundance. In this study, we use MUSE observations to analyze star-forming disc galaxies in 12 clusters within the redshift range \(0.3 < z < 0.5\). Galaxies were classified into three groups: ram-pressure stripping (RPS), control cluster, and control field. For the first time, we investigate the impact of RPS on gas-phase metallicities across a wide mass range of galaxies at intermediate redshift, comparing RPS galaxies to counterparts in both cluster and field environments. By analyzing the integrated flux within galactic disks, our result reveals that, on average, RPS induces a metallicity enhancement of 0.2 dex over non-stripped galaxies. Contrary to the prevailing view that cluster membership alone drives metallicity enrichment, we find that control cluster galaxies exhibit metallicities comparable to field galaxies at a given $\mathrm{M_\star}$, with only RPS galaxies displaying significantly higher metal content, highlighting the unique role of RPS in shaping the chemical properties of galaxies. These differences become more pronounced at lower $\mathrm{M_\star}$, indicating that environmental influences play a more critical role in shaping the chemical evolution of lower-mass galaxies. Our findings suggest that both enhanced star formation rates and suppressed gas inflows—consequences of ram pressure stripping—drive the elevated metallicity observed in RPS galaxies. 

\end{abstract}


\section{Introduction}

The chemical evolution of a galaxy is closely tied to its star formation history (SFH) and the exchange of gas with its surroundings through inflows and outflows \citep[e.g.,][]{2010MNRAS.408.2115M,2012Petropoulou,2019Maiolino,2023Peluso,2024Vulcani,2024Amiri}. In this context, the metallicity of a galaxy serves as an observable indicator, offering insights into how environmental factors influence its SFH and gas content. Much focus has been placed on the mass–metallicity relation (MZR), which highlights the strong correlation between a galaxy's stellar mass and gas-phase oxygen abundance. This relationship has been explored through studies of star-forming galaxies, spanning redshifts $0$ to $\sim 10$ \citep[e.g.,][]{2004ApJ...613..898T,2008ApJ...681.1183K,2008A&A...488..463M,2014ApJ...791..130Z,2015ApJ...799..138S,2020Jones,2023Nakajima}. The shape of the MZR could be influenced by various processes, including variations in the initial mass function (IMF), stellar and active galactic nucleus (AGN) feedback, and the infall of pristine gas \citep[e.g.,][]{2019Maiolino}. The interplay of these factors can shape the MZR and its scatter while providing valuable insights into how environmental conditions impact galaxy properties.

Several potential mechanisms have been suggested for changes in galaxy properties as they migrate from lower-density areas toward the core of galaxy clusters, including intracluster medium (ICM) and interstellar medium (ISM) interactions, the cluster gravitational potential, and minor to major galaxy merging \citep[e.g.,][]{2003Treu,2006Poggianti,2006Boselli,2010Thomas,2023Marasco}. Many previous studies at different redshifts have demonstrated that, for a given mass range, galaxies within clusters exhibit similar or marginally higher metallicities compared to field galaxies \citep[e.g.,][]{2009Ellison,2019Maier,2020Ciocan,2021Sotillo-Ramos,2021Chartab}. In a study of 259 cluster galaxies and 169 field galaxies at redshifts between 0.15 and 0.26, \cite{2019Maier} discovered that cluster galaxies with $\mathrm{\log(M_\star/M_\odot)} < 10.5$ within $\mathrm{R < R_{200}}$\footnote{Referred to as the virial radius, is the radius within which the mean density of a galaxy cluster is 200 times the critical density of the Universe.} exhibited metallicities approximately 0.06 dex higher than the median metallicity of field galaxies. Subsequently, \cite{2020Ciocan} reported an increase in gas-phase metallicity in cluster galaxies (with $\mathrm{9.2 <\log(M_\star/M_\odot)} < 10.2$) by 0.065 dex. Similarly, \cite{2021Sotillo-Ramos} found metallicity differences of approximately 0.08 dex when comparing galaxies in low- and high-density environments. At higher redshifts, studies like \cite{2021Chartab} also observed similar trends, showing consistent patterns in metallicity differences between these environments.

However, several studies have shown that factors other than cluster membership contribute to metallicity enhancements in galaxies. For example, \cite{2009Ellison} found that the slight metallicity increase is primarily influenced by the presence of nearby companion galaxies rather than the cluster environment itself. Similarly, \cite{2024Andrade} speculates that at $z=0.3$, galaxies in clusters not undergoing ram-pressure stripping (RPS) or strangulation exhibit metallicities comparable to or lower than those of field galaxies. In contrast, only galaxies affected by RPS or strangulation are speculated to display elevated metallicities above the MZR of both field and cluster galaxies. As part of the Gas Stripping Phenomena in galaxies with MUSE \citep[GASP,][]{2017ApJ...844...48P} project, \cite{2020Franchetto} demonstrated that cluster galaxies (both stripped and non-stripped) in the local Universe tend to be slightly more metal-rich than field galaxies at  masses below $10^{10.25} \mathrm{M_\odot}$, while exhibiting a similar metallicity trend as field  galaxies at higher stellar masses. Additionally, \citet{2023Peluso} concluded that the gas-phase metallicity in the nuclei of AGN host galaxies is enhanced by approximately 0.05 dex compared to that of star-forming galaxies, irrespective of the presence of RPS.

In this study, building on the work of \cite{2024Khoram}, we utilize MUSE observations to examine the chemical composition of star--forming disc galaxies within 12 clusters at intermediate redshifts, leveraging a statistically significant sample for the first time. We compare the properties of RPS galaxies with those of undisturbed galaxies in both cluster and field environments from the same survey. This comparison aims to identify the dominant environmental factors influencing the chemical evolution of galaxies.

In Sec. \ref{DATA}, we present the galaxy samples used in this study, followed by Sec. \ref{data analysis}, which outlines the details of the spectral analysis and the methods used to measure oxygen abundance in the sample. Our findings are presented in Sec.\ref{RESULT} and discussed further in Sec.\ref{DISCUSSION}.

\section{Data and sample selection}\label{DATA}

We utilize MUSE observations acquired through the MUSE GTO project \citep{2021Richard}, which targeted clusters selected from various surveys including the Massive Clusters Survey \citep{2001Ebeling}, Frontier Fields \citep{2017Lotz}, GLASS \citep{2015Treu}, and the Cluster Lensing and Supernova survey with Hubble \citep{2012Postman} programs. Observations were confined to the central regions of the clusters, with the typical area covered by MUSE observations corresponding to around 250–330 kpc, contingent on the cluster's redshift. This corresponds to roughly the inner 0.1–0.15 $R_{\rm 200}$. The clusters have masses in the range of $\sim 10^{13.3} - 10^{14.4} M_\odot$ and velocity dispersions between $\sim 800$ and $1800$ km/s \cite{2021Richard}. These clusters were observed either through single pointings or mosaics, with exposure durations ranging from approximately 2 to 15 hours (effective). For further insights into the observations and data analysis, we refer to \cite{2021Richard}.

As outlined in \cite{2022ApJ...925....4M}, \cite{2022Werle}, and \cite{2024Vulcani}, we selected a sample of 12 clusters\footnote{Abell2744, Abell370, AS1063NE, AS1063SW, MACS0257, MACS0416N, MACS0416S, MACS0940, MACS1206, RXJ1347, SMACS2031, SMACS2131} within the redshift range $0.3 < \text{z} < 0.5$. This sample encompasses star--forming disc galaxies spanning a stellar mass range of $7.8 \lesssim \log(\mathrm{M_\star/M_\odot}) \lesssim 11.8$. We then classified galaxies into three groups: (1) Galaxies undergoing RPS are identified by examining extraplanar, unilateral tails, and/or debris with emission lines in the MUSE data cubes and/or HST images, (2) Galaxies belonging to the cluster but not experiencing ram-pressure (Control Cluster, or CC) and (3) Galaxies in coeval fields non members of any cluster (Control Field, or CF). The initial sample contains 28 RPS, 40 CC, and 12 CF galaxies in total (Moretti et al. in prep). Notably, the H$\alpha$ emission line falls beyond the MUSE wavelength range for galaxies with redshift $\gtrsim 0.42$. Consequently, 12 RPS, 12 CC, and 4 CF galaxies lack H$\alpha$, [S II]$\lambda\lambda6718,6730 \AA$, and [N II]$\lambda\lambda6548,6584 \AA$ emission lines. We will keep these galaxies separate for the rest of the analysis and introduce a new calibration method to estimate their metallicities based on \Oii\ and \Oiii\ fluxes. 

Through statistical analysis and visual inspection of individual galaxy disks, we excluded flawed, contaminated, or faint galaxies from the sample. For example, three RPS galaxies were determined not to be cluster members after phase-space investigations based on the \Ha redshift. Additionally, in the CC sample, three galaxy disks were found to be contaminated by outflows and/or AGN flux. To ensure robust metallicity estimation, we included only galaxies in which at least two-thirds of the disk spaxels have a flux with signal to noise (S/N) $> 2$ for each emission line used in each metallicity diagnostic (see Sec.\ref{data analysis}). We further excluded interacting galaxies to minimize sources of contamination. Ultimately, we confidently selected 18 RPS, 23 CC, and 9 CF galaxies for disk metallicity measurements. We confirm that both RPS and control galaxies are represented in most clusters, ensuring that our results are not influenced by systematic environmental differences between the two classes.

\section{Data Analysis}\label{data analysis}

First, we correct the MUSE spectra for the Milky Way dust extinction, as extensively discussed in \cite{2022ApJ...925....4M}. The data cubes are then spatially smoothed using a boxcar kernel of 5×5 pixels (approximately 1 arcsecond, corresponding to the average seeing of the cubes). From these smoothed cubes, we extract the stellar-only component by fitting the spectrum within each spaxel using the SINOPSIS code \citep{2017ApJ...848..132F} along with the latest stellar population models from \cite{2003Bruzual}, assuming a \cite{2003PASP..115..763C} initial mass function (IMF). The emission-only spectrum is then obtained by subtracting the fitted stellar component. Gas kinematics, emission line fluxes, and the respective errors are derived utilizing HIGHELF software (Radovich et al. 2025, in preparation), built upon the LMFIT Python package\footnote{https://lmfit.github.io/lmfit-py/}.

 In this work, to delineate the boundaries of each galaxy, we create an SDSS g-band image \citep[Sloan Digital Sky Survey,][]{SDSS} by conducting synthetic photometry on the MUSE data cube. Using this image, we generate masks to identify pixels with flux values exceeding the background level by 5$\sigma$. This approach was initially introduced in \cite{2022ApJ...925....4M} and further elaborated on in \cite{2024Khoram}.

Spaxels within the galactic disks are treated differently based on the presence or absence of H$\alpha$ within the observed spectral range, depending on their redshifts.

\subsection{With \Ha}

 When H$\alpha$ is contained in the galaxy spectral range (68\%/70\%/89\%  of the  RPS,  CC, and CF  samples, respectively), we use the BPT diagram \citep{1981PASP...93....5B} and the O3--N2 demarcation line by \cite{2001ApJ...556..121K} to exclude non-star-forming spaxels.  89\% of the spaxels in the control sample (i.e. both CC and CF) and 85\% of spaxels in RPS galaxies are star-forming within the disks. Fluxes are then corrected for intrinsic dust extinction using the \cite{1989ApJ...345..245C} extinction curve assuming an intrinsic Balmer decrement of $\mathrm{H\alpha/H\beta } = 2.86$. Moreover, for galaxies with \Ha and the \Nii emission lines, we derive the metallicity using the calibration introduced by \citeauthor{2004MNRAS.348L..59P} (\citeyear{2004MNRAS.348L..59P}, hereafter P04) according to the formula $\mathrm{ 12 + \log (O/H) = 8.73 - 0.32 \times O3N2}$. This is an empirical method, calibrated via electron temperature measurements, that determines the oxygen abundance in \hii regions based on the O3N2 index, defined as $\mathrm{\log[ (\Oiii/H\beta)/(\Nii/H\alpha)]}$. This method is applicable when O3N2 is less than 1.9, corresponding to a minimum metallicity of approximately 8.1. Note that $3\times[\text{N}\,\textsc{ii}]\lambda6548$ flux is employed as substitute to \Nii in 3 cases where the latter line falls beyond the covered wavelength range. 

\begin{figure}
    \centering
    \includegraphics[width=\linewidth,height=\textheight,keepaspectratio]{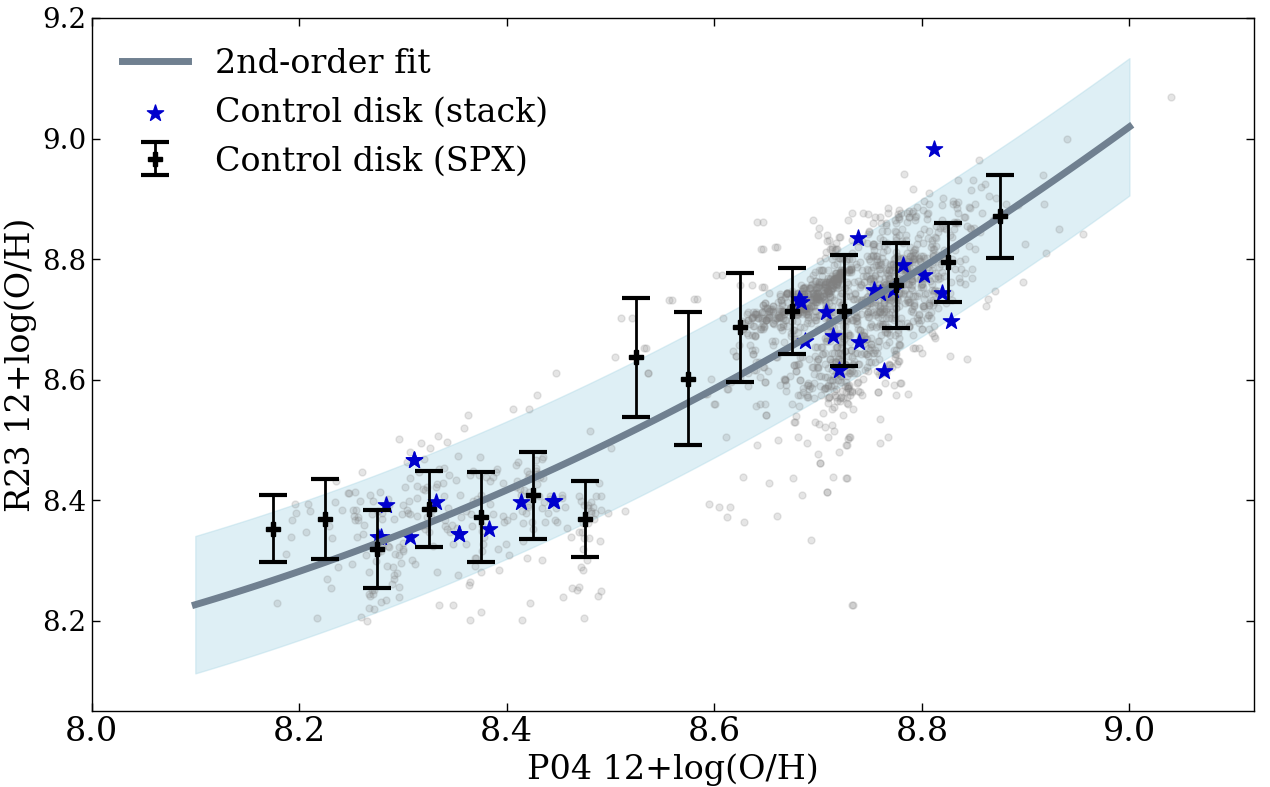}
    \caption{Metallicity values based on the R23 index from \citeauthor{2021Sanders} (\citeyear{2021Sanders}, S21) and the O3N2 index from \citeauthor{2004MNRAS.348L..59P} (\citeyear{2004MNRAS.348L..59P}, P04), superimposed with the second-order relation derived from all star-forming spaxels (gray) in the control sample. The standard deviation of data points from the fit (0.09 dex) is illustrated with the shaded region. Black data points denote the median S21 spaxel metallicities in 0.02 dex P04 bins, with error bars representing the $1\sigma$ percentiles for visual clarity. Blue stars represent the global (i.e. integrated) disk metallicities of the control sample. }
    \label{fig:R23_Rel}
\end{figure}

\begin{figure*}
    \centering
    \includegraphics[width=\linewidth,height=\textheight,keepaspectratio]{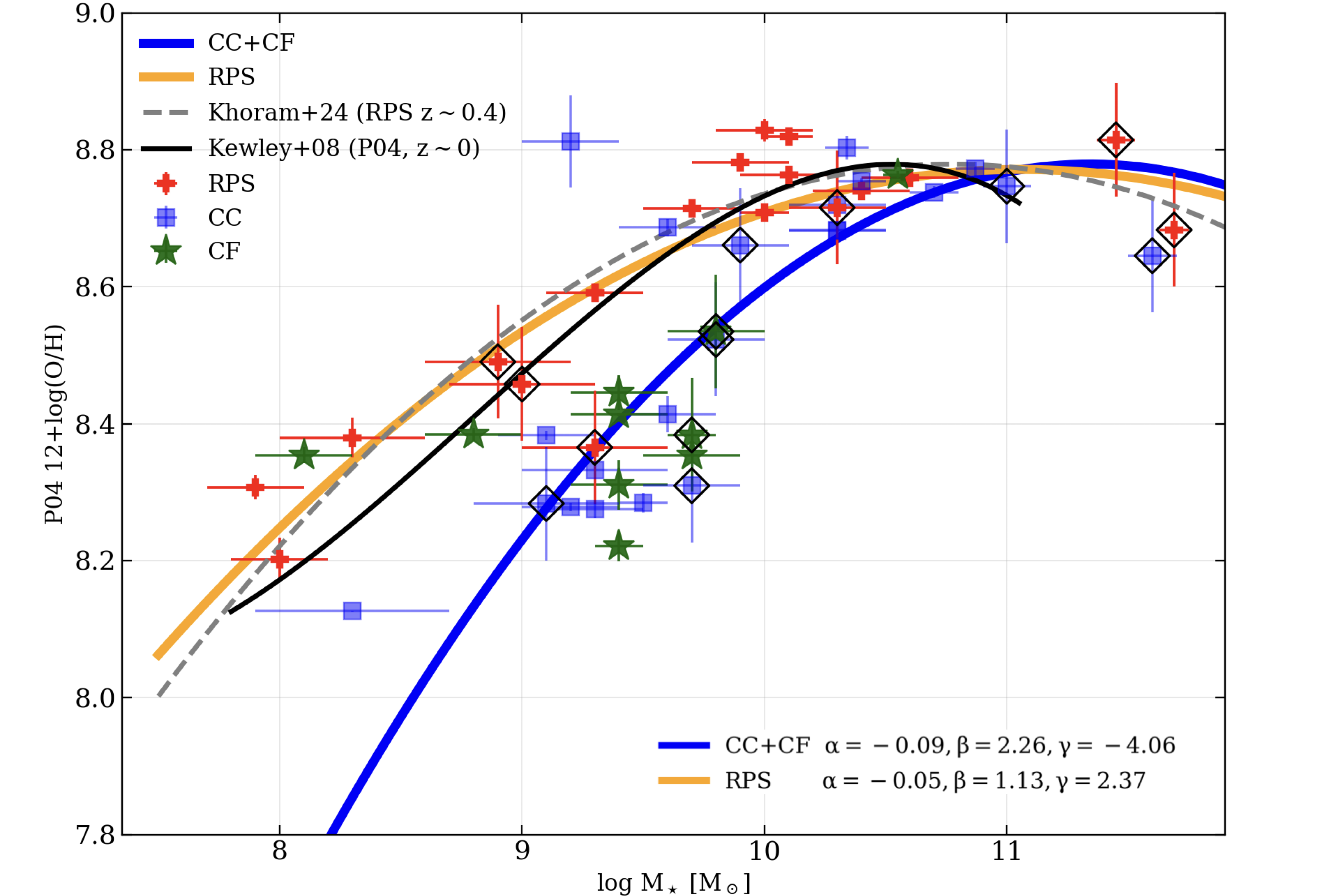}
    
    \caption{Measured global MZR for all galaxies in the sample from integrated disk spectra (see Sec.\ref{data analysis}). The orange line represents an error--weighted second--order polynomial fit to RPS galaxies (red plus symbols), while the blue line shows the same fit to CC and CF galaxies, depicted by blue squares and green stars, respectively. For comparison, the dashed line shows the fit from \citet{2024Khoram} for RPS galaxies in the Abell 370 and Abell 2744 clusters. The solid black line represents the MZR derived by \citet{2008ApJ...681.1183K} employing the P04 metallicity calibration and \citet{2003PASP..115..763C} IMF. Data points with superimposed diamonds mark galaxies where \Ha falls outside their wavelength coverage, and their metallicities are derived from \Oii\,  fluxes (see Sec.\ref{data analysis}). The fit parameters are displayed in the bottom-right of the figure, corresponding to the coefficients in the general second-order equation ($\mathrm{12+log(O/H) = \alpha \cdot logM^2_\star + \beta \cdot logM_\star + \gamma}$).}
    
    \label{fig:MZR}
\end{figure*}

\subsection{No \Ha}
% \noindentNo \Ha}

For galaxies with \Ha outside the observed spectral range (32\%/ 30\%11\% of the RPS, CC, and  CF samples, respectively), we use a relation based on the R23 index, defined as \Oiii + \Oii)/\Hb, and its correlation with metallicity introduced in \cite{2021Sanders}. Using all the star-forming spaxels from the control sample (CC and CF), we establish a second-order relation between the R23-based metallicity and that derived using P04 ($\mathrm{Z_{P04}}$) as given in Eq.\ref{eq: R23_P04 met} with the standard deviation of $0.09 \mathrm{dex}$ from the polynomial fit. 
\begin{equation}\label{eq: R23_P04 met}
    12+\mathrm{log(O/H)_{R23}} = 0.18 \times \mathrm{Z^2_{\textsc{P04}}}  -2.34 \times \mathrm{Z_{\textsc{P04}}}  + 15.21
\end{equation}

As illustrated in Fig.\ref{fig:R23_Rel}, $95\%$ of the disk (integrated flux) metallicity values fall within the uncertainty range of this relation. Consequently, with the \Oiii, \Oii, and \Hb fluxes, we can reliably estimate P04 metallicities for objects whose \Ha line falls outside the MUSE wavelength range, with a reasonable level of confidence. As detailed in Moretti et al.\ (in preparation), we estimate dust extinction in the absence of the Balmer decrement using an empirical relation based on the [O\,II]/H$\beta$ ratio for massive galaxies (\(\log M > 10\)). For lower-mass galaxies, we assume a constant \(A_V = 1\).

In both cases, whether \Ha is present in the spectral coverage or not, metallicity uncertainties are computed using Monte Carlo simulations, where all recorded line fluxes are randomly perturbed 1000 times according to their measurement errors, assuming a Gaussian noise distribution. The resulting metallicities are taken as the median of these simulations, with the reported uncertainties given by their standard deviations. However, it is important to note that these values likely underestimate the true uncertainties, as systematic contributions are not accounted for.

\section{Result}\label{RESULT}

\begin{figure}
    \centering
    \includegraphics[width=\linewidth,height=\textheight,keepaspectratio]{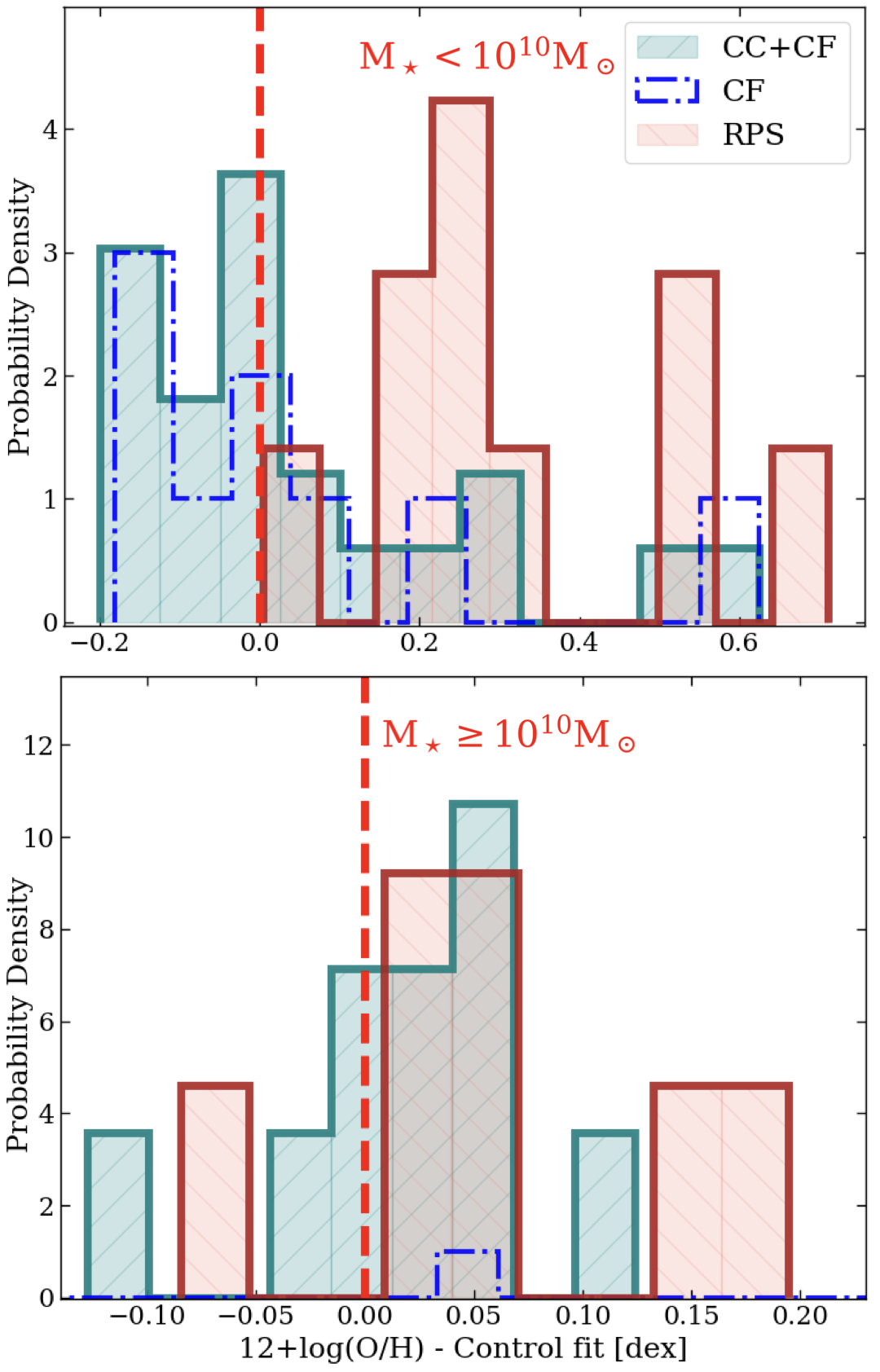}
    \caption{The normalized metallicity residuals of RPS and control galaxies, relative to the control (CC+CF) fit, are shown in Fig.\ref{fig:MZR}. Galaxies are categorized into two groups based on their stellar mass, below and above $\mathrm{10^{10}M_\odot}$, to emphasize the distinct behavior across different mass ranges. The green histogram includes CC and CF collectively, while the blue one demonstrates only CF residuals. The red histogram shows RPS residuals from CC+CF fit (dashes red line).} 
    \label{fig:Hist_MZR}
\end{figure}

We derive the disk metallicity for each galaxy from their integrated spectra, following the criteria outlined in Sec.\ref{data analysis}. As illustrated in Fig.\ref{fig:MZR}, both galaxies with and without \Ha in their spectral range (see Sec.\ref{data analysis}) are included, with those lacking \Ha indicated by a diamond superimposed on their data points. We compare the disk's gas--phase metallicity of RPS, CC, and CF galaxies as a function of their stellar mass. We provide MZRs (error--weighted second--order polynomial fit) for our sample, spanning the stellar mass from $10^{8}$ to $10^{12} \, \mathrm{M_\odot}$. 

In Fig.\ref{fig:MZR}, the orange line corresponds to the mass--metallicity relation of RPS galaxies, represented by red plus symbols, while the blue line indicates the fit to the CC and CF galaxies, shown as blue squares and green stars, respectively. We confirm that excluding CF galaxies from the blue fit does not affect the control MZR, as CC and CF galaxies with similar stellar masses exhibit comparable metallicity values. Furthermore, we compared our results to the MZR derived by \citet{2008ApJ...681.1183K}, as shown in Fig.~\ref{fig:MZR}. Their analysis was based on a sample of 27,730 star-forming galaxies with \(z < 0.1\) from the SDSS DR4 catalog, using fiber-integrated spectra. The study originally employed the P04 metallicity calibration and the \citet{1955ApJ...121..161S} IMF, which we converted to the \citet{2003PASP..115..763C} IMF for consistency with our analysis.

As illustrated in Fig.\ref{fig:MZR}, a meaningful comparison between RPS and the control sample with $\mathrm{log(M_\star/M_\odot)} < 9$ is limited due to the small number of CC and CF galaxies in this range. However, in the mid-- to high--mass regimes, RPS galaxies exhibit noticeably higher metallicities compared to galaxies of similar mass in the control sample. The distinction becomes more apparent in Fig.\ref{fig:Hist_MZR}, which represents the metallicity residuals for the entire sample relative to the CC+CF MZR (the blue line in Fig.\ref{fig:MZR}), separately for galaxies with stellar masses below and above $\mathrm{10^{10}M_\odot}$. The median metallicity residual for RPS galaxies is approximately 0.2 dex across the entire mass range. However, the most significant deviation, about 0.3 dex, is observed in the low- to mid-mass range around $\mathrm{log(M_\star/M_\odot)} \sim 9.5$, with this difference gradually diminishing at higher masses.

This result suggests that environmental influences have a more significant impact on lower and mid--stellar mass galaxies compared to their higher--mass counterparts.

\section{Discussion}\label{DISCUSSION}

Although stellar mass is the primary factor driving chemical enrichment in galaxies, other physical conditions also play a role in their chemical evolution \citep[e.g.,][]{2019Maiolino}. A number of studies have demonstrated that environmental factors can affect the metal enrichment of galaxies within similar stellar mass ranges \citep[e.g.,][]{2019Maier,2020Ciocan,2021Sotillo-Ramos,2021Chartab,2024Andrade}. Our results indicate that, regardless of cluster membership, ram-pressure stripping alters significantly the metal content in galaxies' gas (see Figures \ref{fig:MZR} and \ref{fig:Hist_MZR}). 

Previous studies \citep[e.g.,][]{2014PengMaiolino,2016Maier,2019Maier,2020Ciocan} have shown that cluster membership alone can lead to a metallicity enhancement of approximately \(0.06{-}0.1\) dex compared to field galaxies at \(z \simeq 0{-}0.4\). These findings are based on comparisons between cluster and field galaxy populations. The primary explanation proposed for this effect was strangulation (or starvation), a mechanism that depletes the diffuse hot gas reservoir within the galaxy's halo, while leaving the disk's gas largely unaffected. Additionally, \cite{2014PengMaiolino} introduced the idea that gas inflow into the halo of a galaxy within a dense cluster environment becomes progressively enriched with metals. This enrichment leads to higher metallicities in low-mass cluster galaxies compared to similar galaxies in less dense, field environments. However, they did not propose a specific mechanism by which this metal-enriched gas inflow persists in dense galaxy clusters. Recent studies \citep[][]{2020Ciocan,2024Andrade} have speculated that, in addition to strangulation, RPS may also contribute to this enhancement. However, they did not provide detailed analysis or further exploration of this idea.

 To our knowledge, only one study, \cite{2020Franchetto}, has addressed the presence of RPS galaxies in the cluster sample and distinguished them from other galaxy types. The authors demonstrated that, for a given mass, cluster and field galaxies at $\mathrm{z\simeq 0}$ show only minor differences in metallicity at 1Re, which are not statistically significant. At the higher mass end, RPS galaxies also show comparable levels of metal enrichment, comparable to our findings. While their sample includes a few low- and intermediate-mass RPS galaxies that show higher metallicity compared to both cluster and field samples, the authors refrained from drawing robust conclusions about these differences due to the limited sample size at $\mathrm{log( M_\star/M_\odot) \lesssim 10.4}$. In addition, both numerical simulations  and analytical models predict that the effectiveness of RPS increases with redshift, driven by its dependence on the density of the ICM \citep{2012Kravtsov,2019Singh,2019Mostoghiu}, implying a heightened impact of RPS in the redshift range of our sample.

One could argue that, due to having composite and/or low S/N spaxels in the outer parts of RPS galaxies \citep[e.g.,][]{2022ApJ...925....4M}, global metallicity values predominantly reflect the more central regions of these galaxies, with any enhancements originating there. However, \cite{2024Khoram} have shown that the metallicity profile of RPS galaxy disks at intermediate redshifts remains flat or slightly positive in a range of  $\mathrm{M_\star < 10^{10}\,M_\odot}$, where the largest differences between RPS and control metallicities are observed. Additionally, various studies \citep[e.g.,][]{2017MNRAS.469..151B,2018MNRAS.478.4293C,2024Khoram_Direct_met} have shown that galaxies with $\mathrm{M_\star < 10^{9.5}\,M_\odot}$ at local to intermediate redshifts exhibit relatively flat metallicity gradients, suggesting that the average global metallicity remains unchanged even when the outermost disk regions are excluded. This stability supports the idea that a physical mechanism may underlie the observed enhancement in RPS metallicities.

Considering the metal content of galaxies as a balance between stellar evolution over their lifetime and the inflow of cold, metal-poor gas, the metallicity enhancement in RPS galaxies could be explained by the influence of ram-pressure on these two factors. Compared to typical star-forming galaxies in both clusters and the field, RPS galaxies display elevated SFR on both global and local scales, likely driven by processes such as gas compression and mass flows induced by ram-pressure \citep[e.g.,][]{2016AJ....151...78P,2018ApJ...852...94V,2020Vulcani_B,2023ApJ...950...24B,2023Giunchi}. By analyzing the full sample of 12 clusters in this study and applying the same BPT diagram criteria to identify star-forming spaxels for SFR estimation, \cite{2024Vulcani} observed a mild SFR enhancement in RPS galaxies with ionized gas tails compared to the control sample (non-RPS galaxies). Enhanced SFR in RPS galaxies were found also in lower redshift clusters \citep[][]{2018ApJ...852...94V,2020Roberts}. On the global galactic scale, the SFR enhancement is mild, while, the analysis of post-starburst features in the spectra of galaxies in our clusters \citep{2022Werle} has shown that locally, wherever in the disk there is gas left, the burst of star formation is strong and forms a significant fraction of the local mass (see Fig.5 in \citealt{2022Werle}). Since the measured metallicity reflects the gas that remains in the galaxy, the SFR increase can significantly influence the gas metallicity observed in RPS galaxies. 

Note that the RPS signature we observe, characterized by long tails of ionized gas, typically becomes visible a few hundred Myr after ram-pressure begins to affect an infalling galaxy \citep{2022Smith,2024Werle}. Prior to this, the galaxy's surrounding hot gas halo is rapidly stripped by RPS \citep{2023Kulier}. Additionally, the chemical enrichment driven by RPS-enhanced star formation occurs on short timescales, governed by the lifetimes of the most massive stars. The interplay of these effects likely explains the measurable chemical enrichment and the mildly enhanced SFR observed in these galaxies.

In addition, some studies \citep[e.g.,][]{2013Hughes,2015Peng} have demonstrated that RPS is responsible for cutting off the inflow of the cold metal-poor gas to the central regions. Consequently, it is possible for gas inflow to be halted while the galaxy continues forming stars, resulting in a metallicity enhancement on short timescales. In this scenario, the metallicity enhancement observed in RPS galaxies results from a combination of two key factors: (1) increased metal production driven by the locally enhanced SFR, and (2) the suppression of inflow of low-metallicity gas caused by ram-pressure stripping. Simulations further show that these effects reinforce each other to some extent. Using results from the EAGLE simulation \citep{2015Schaye}, \citet{2023Kulier} demonstrated that stellar feedback enhances the efficiency of ram-pressure stripping. This interplay indicates that the SFR enhancement associated with RPS not only amplifies gas stripping but also inhibits the inflow of cold, metal-poor gas, thereby altering the metallicity and chemical enrichment of RPS galaxies.

We confirm that the observed enhancement in SFR among metal-rich RPS galaxies, compared to normal galaxies unaffected by stripping, stands in clear contrast to the hotly-debated Fundamental Metallicity Relation \citep[e.g.,][]{2010MNRAS.408.2115M,2019Maiolino,2020MNRAS.491..944C,2024Khoram_Direct_met}. This finding further supports the notion that environmental effects are the primary drivers of the differences observed in Fig. 2 between the CC+CF and RPS samples.

From a different perspective, the MZR evolution demonstrates a clear and consistent evolution with redshift, where the metallicity decreases at a fixed mass as the redshift increases. At lower redshifts, this evolution is more pronounced in lower-mass galaxies, whereas higher-mass galaxies seem to have achieved their present-day metallicity by approximately $z \sim 1$ \citep[see][and references therein]{2013ApJ...771L..19Z,2019Maiolino}. As illustrated in Fig.~\ref{fig:MZR}, the metallicities of RPS galaxies remain closely aligned with the \cite{2008ApJ...681.1183K} MZR at \(z \sim 0\), represented by the solid black line. In contrast, the control sample exhibits lower metallicities, as expected, highlighting an overall enhancement in the metallicity of RPS galaxies compared to the average values predicted for their redshift.

In summary, our findings show a clear distinction between RPS and non-RPS cluster galaxies, the former having higher metallicities than the latter, with no-RPS galaxies sharing the same mass-metallicity relation of field galaxies. This result favors the scenario that cluster membership solely cannot be responsible for experiencing different chemical evolution and metal content of galaxies within the similar stellar mass range. We also challenge the scenario which introduces strangulation (or starvation) as the primary explanation proposed for this effect \citep[e.g.,][]{2016Maier,2019Maier,2020Ciocan}. It is important to note that, given the relatively small size of our sample compared to studies that have derived the MZR at local and intermediate redshifts, further investigation is required to robustly confirm the findings presented in this work.

\section{Summary and Conclusions}

In this study, we analyzed a total sample of 50 star--forming disc galaxies to investigate the impact of RPS on the gas-phase metallicity of galaxies across 12 clusters in the redshift range \(0.3 < z < 0.5\). Using MUSE spectroscopy, we measured the gas-metallicity of galaxy disks and compared RPS galaxies to CC and CF galaxies. Our analysis yielded several key findings:

\begin{enumerate}
    \item  
    The global MZR for RPS galaxies shows a systematic metallicity enhancement compared to the control sample, particularly in the low- to intermediate-mass regime (\(\mathrm{log(M_\star/M_\odot)} \sim 8.0\) to \(9.5\)). This enhancement diminishes at higher stellar masses, suggesting that environmental effects are more significant in shaping the chemical evolution of lower-mass galaxies. Metallicity residuals relative to the control MZR confirm that RPS galaxies exhibit elevated metallicities, with the median deviation of approximately 0.2 dex observed in the mid-mass range. This result indicates that RPS significantly alters the chemical enrichment of galaxies beyond the typical effects associated with cluster membership alone.

    \item The metallicity enhancement in RPS galaxies can be attributed to a combination of two primary mechanisms:
    \begin{enumerate}
        \item Enhanced SFR: Locally elevated SFRs in RPS galaxies, driven by gas compression and mass flows due to ram-pressure, boost metal production in the remaining gas. This is supported by our observation of locally intense star formation bursts in RPS regions.  
        \item Suppressed Gas Inflows: RPS cuts off the supply of metal-poor gas to the central regions, preventing dilution and allowing metal-enriched gas to dominate the disk's chemical composition.  
    \end{enumerate}

    \item While previous studies suggest that cluster membership alone enhances metallicities by \(0.06-0.1\) dex compared to field counterparts, our results demonstrate that RPS can lead to significant enrichment. This finding underscores the complex interplay between SFR enhancement, gas dynamics, and environmental stripping processes. Furthermore, theoretical models suggest that feedback from enhanced star formation not only amplifies stripping efficiency but also suppresses gas inflows, further reinforcing the observed metallicity trends.

\end{enumerate}

Our findings provide robust evidence that RPS plays a crucial role in shaping the chemical evolution of galaxies, particularly in dense cluster environments. The observed metallicity enhancement in RPS galaxies underscores the importance of environmental effects in galaxy evolution and offers a deeper understanding of how stripping processes and star formation interact to regulate the metal content of gas in galaxy disks. Future work combining high-resolution simulations and spatially resolved observations will further elucidate the detailed mechanisms driving these trends.

\section*{Acknowledgements}

This project, funded by the European Research Council (ERC) under the European Union’s Horizon 2020 research and innovation programme (grant agreement No. 833824), is based on observations gathered at the European Organization for Astronomical Research in the Southern Hemisphere, under the ESO programme 196.B-0578. This work used products from HFF-DeepSpace, funded by the National Science Foundation and Space Telescope Science Institute (operated by the Association of Universities for Research in Astronomy, Inc., under
NASA contract NAS5-26555).

\section*{DATA AVAILABILITY}

The data underlying this article will be shared on reasonable
request to the corresponding author.
% \begin{acknowledgement}
% \end{acknowledgement}


\bibliographystyle{mnras}
\bibliography{ref}

%\begin{figure*}
    % \centering
    % \includegraphics[width=\linewidth,height=\textheight,keepaspectratio]{Figure/SFR_Coded.png}
    % \caption{Caption}
    % \label{fig:enter-label}
%\end{figure*}
%%%%%%%%%%%%%%%%%%%%%%%%%%%%%%%%%%%%%%%%%%%%%%%%%%

% Don't change these lines
% \bsp	% typesetting comment
\label{lastpage}
\end{document}